\documentclass[twocolumn,preprintnumbers,amsmath,amssymb,pre]{revtex4}

\usepackage{graphicx}
\usepackage{dcolumn}
\usepackage{bm}

\begin{document}

\title{Analysis of the Airport Network of India as a complex weighted network}
\author{Ganesh Bagler}
\email{ganesh@ncbs.res.in}
\affiliation{National Centre for Biological Sciences, Tata Institute of Fundamental Research, Bangalore 560065, India.\footnote{The author was formerly at Centre for Cellular and Molecular Biology, Uppal Road, Hyderabad 500007, India.}}
\date{\today}

\begin{abstract}
Transportation infrastructure of a country is one of the most important indicators of its economic growth.
Here we study the Airport Network of India (ANI), which represents India's domestic civil aviation infrastructure,
as a complex network. 
We find that ANI, a network of domestic airports connected by air links, is a small-world network characterized by a 
truncated power-law degree distribution, and has a signature of hierarchy.
We investigate ANI as a weighted network to explore its various properties and compare them with their topological counterparts.
The traffic in ANI, as in the World-wide Airport Network (WAN), is found to be accumulated on interconnected groups of airports and 
is concentrated between large airports. 
In contrast to WAN, ANI is found to be having disassortative mixing which is offset by the traffic dynamics.
The analysis indicates toward possible mechanism of formation of a national transportation network, which is different from that on 
a global scale.
\end{abstract}


\maketitle


\section{\label{sec:intro}Introduction}
Transportation infrastructures are of crucial importance to the development of a country~\cite{raghuraman1998} and are 
important indicators of its economic growth. 
They form the backbone of tourism industry, support movement of goods and people across the country, thereby driving the national economy.
Roadways, railways, and airways are the major means of transport in India, although contribution of airways is small compared 
to that of the other two. 
The civil aviation sector in India has been developing steadily and is expected to grow at an accelerated pace with the shift in 
the policy of the government, and the addition of several low-cost private air service providers~\cite{report:bw}. 
These private players offer competitive pricing, provide more region-oriented service, and thereby help to increase the air traffic.
Understanding of these transportation systems is important for reasons of policy, administration and efficiency.

During the past few years, complex network analysis has been used to study 
transportation systems (railways~\cite{transport:manna}, 
airlines~\cite{air-amaral-PNAS,air-WAN,air-china,air-WAN-PNAS,air-WAN-model}), 
which are man-made infrastructures, 
from different aspects, apart from many other systems of diverse origins~\cite{reka:thesis,dorogovtsev:book}.
The World-wide Airport Network (WAN) has been studied from topological as well as traffic dynamics perspective. 
WAN has been studied~\cite{air-amaral-PNAS} for its degree distribution and a model with constraints (such as cost of adding 
links to the nodes or the limited capacity of nodes) was proposed to account for the truncation for high-degrees in its scale-free 
cumulative degree distribution. 
It has been observed~\cite{air-WAN} that in WAN, most connected nodes are not necessarily the most ``central'' -- nodes through 
which most of the shortest paths go.
A model, incorporating the geo-political constraints, has been proposed~\cite{air-WAN-model} to explain this 
apparently surprising result.
Beyond the topological features, WAN was studied~\cite{air-WAN-PNAS} in more detail by considering the traffic -- strength of 
interactions between nodes -- dynamics on it.
A model was proposed~\cite{barrat:PRL} for the evolution of weighted evolving networks in an effort to understand the statistical 
properties of real-world systems.
Airport Network of China (ANC)~\cite{air-china}, a network much smaller than WAN, was also analyzed for its topology 
and traffic dynamics. Its topology was found to be having small-world network features and a two-regime power-law degree 
distribution. 

We investigate India's domestic airport network which comprises air services of all major civil air service providers. 
We study the network for its topological features and for its traffic dynamics, by considering the intensity of interactions.
First, we study the network as an ``unweighted network'' to investigate its topological properties. 
This network is formed by considering whether or not a pair of airports are connected by an airline. 
As in many other complex networks, the details of flow of information is a crucial factor for airport networks. 
Hence we study ``weighted network'', in which we consider the number of flights plying between any two airports per week. 
``Unweighted network'' represents the architecture (topology) of airports' connectivity, whereas ``weighted network'' represents 
traffic dynamics in the network. 

Our analysis shows that while ANI is similar to the WAN in some aspects, it has differences in some features as reflected in 
its network parameters.
We find that ANI has small-world~\cite{watts:book} network features and has a scale-free~\cite{SFNW:barabasi} degree distribution. 
The traffic in ANI is found to be accumulated on interconnected groups of airports and is concentrated on trunk routes between 
large airports. It is found to be having a disassortative topology, in contrast to WAN. 

\section{\label{sec:ani}Airport Network of India (ANI)}
The Airport Network of India (ANI) comprises domestic airports of India and airlines connecting them.
The air services provided in ANI include all major domestic air service providers in India, including some international airlines 
which ply domestic routes. The data was obtained from a timetable 
(EXCEL's Timetable for air services within India~\footnote{The timetable contains a few airports ($6$) of neighboring countries in the 
close vicinity of India.}, as on 12th January 2004) of air services, which includes 
domestic services provided by following air-travel service providers: Indian Airlines, Alliance Air, Jet Airways, Air Sahara, 
Air Deccan, Jagson, Druk Air, Air India, Bangladesh Biman, Royal Nepal and Srilankan Airlines.

\subsection{Unweighted ANI}
The ``unweighted ANI'' is a directed network with ($N$=) 79  nodes (airports) and 442  
directed links (flights going from one airport to another). It is represented by a binary adjacency matrix, $A$($N \times N$), 
whose elements $a_{ij}$ take value $1$ if there is a flight by any service provider from airport $i$ to airport $j$ on any day of 
the week and $0$ otherwise. 
The asymmetric adjacency matrix ($A$) was used to find properties (in-degrees, out-degrees, and shortest paths) which are sensitive
to the direction.
Certain properties (degrees, degree-correlations, clustering coefficients) were calculated after symmetrizing the adjacency matrix, 
which is justified, as out of total $228$ flight routes $221$ are bi-directional. We needed to add only 14 fictitious flights 
so as to symmetrize $A$. The total number of edges in the symmetrized adjacency matrix is, $M=228.$

\subsection{Weighted ANI}
To include the information about the amount of traffic flowing on the network, the ``weighted ANI'' is defined 
by considering the strengths of the links in terms of number of flights per week.
It is represented by a weight matrix, $W$, where each element $w_{ij}$ stands for the the total number of flights per week from 
airport $i$ to airport $j$. Since majority of the nodes are symmetric in ANI, in terms of traffic flowing into and 
out of it ($w_{ij}=w_{ji}$), we symmetrize the weight matrix $W$ which is used for all the weighted analyses.

\section{\label{sec:uw-ani}Topological Analysis of Unweighted ANI}
Degree of a node is the number of nodes it is directly connected to. Degree of a node $i$ is defined as,
\begin{equation}
k_{i}=\sum_{j=1}^{N} a_{ij}.
\label{eq:degree}
\end{equation}
In a directed network, in-degree (out-degree) of a node is the number of in-coming (out-going) links. 
In ANI, in-degree ($k_i^{in}$) and out-degree ($k_i^{out}$) of an airport stand for the number of 
flights terminating-into and number of flights originating-from that airport, respectively. 
We observe that for a very large number of nodes in ANI $k_i^{in} =  k_i^{out}.$
The average degree of symmetrized ANI was found to be $\langle K \rangle=2M/N=5.77.$
The average shortest path length ($L$) for a directed network with $N$ nodes is defined as,
\begin{equation}
L = \frac{1}{{N(N-1)}} {\sum_{\substack{i,j=1\\i \ne j}}^N L_{ij}}, 
\label{eq:avg_path_length}
\end{equation}
where $L_{ij} \equiv$ shortest path length from node $i$ to $j$.
Clustering coefficient ($C_i$) of a node is defined as the ratio of number of links shared by its neighboring nodes to the
maximum number of possible links among them. In other words, $C_i$ is the probability that two nodes are linked
to each other given that they are both connected to $i$.
The average clustering coefficient is defined as,
\begin{equation}
C =\frac{1}{N} {\sum_{i=1}^N C_{i}},
\label{eq:avg_cc}
\end{equation}
where $C_i \equiv$ clustering coefficient of node $i$.

Small-world~\cite{watts:nature,watts:book} networks are characterized by a very small average shortest path length ($L$)
and a high average clustering coefficient ($C$). Shortest path length from node $i$ to $j$, $L_{ij}$, is the number of flights 
needed to be taken to go from $i$ to $j$ by the shortest route.
We found the average shortest path length of ANI to be ($L$=) $2.2593$, which is of the order of that of a random network 
($L_{rand} \sim \ln{N}/\ln{\langle K \rangle} = 2.493$) of same size and average degree. 
The average clustering coefficient of ANI was found to be ($C$=) $0.6574$, which is an order of magnitude higher than that of 
the comparable random network ($C_{rand} \sim \langle K \rangle /N = 0.0731$). These two properties indicate that ANI is a 
small-world network. WAN~\cite{air-WAN,air-WAN-PNAS} and ANC~\cite{air-china} have also been found to be small-world networks.
In particular, ANC, which is of comparable size of that of ANI, has average shortest path length and clustering coefficient, $2.067$
and $0.733$, respectively. 

\begin{figure}
\includegraphics{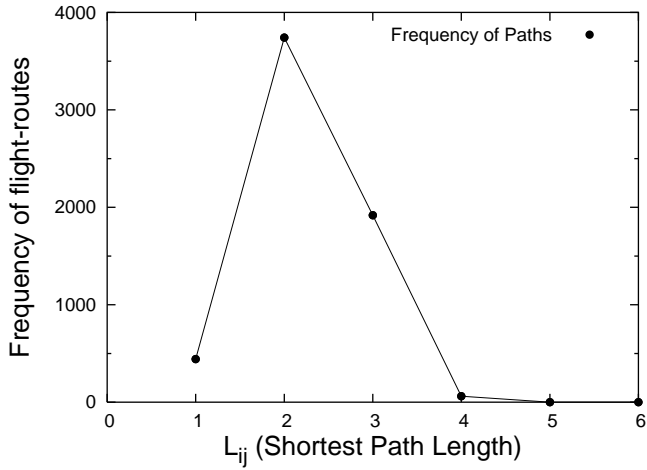}
\caption{\label{fig:shortest_path_dist}Shortest path distribution in ANI}
\end{figure}

\subsection{\label{subsec:shortest_path_dist}Shortest Paths Analysis}
ANI (a directed connected network with $79$ nodes) has $6162$  distinct paths -- node-to-node flight routes. 
Shortest path analysis is intended to give us an idea about the ease of travel in the network.
Fig.~\ref{fig:shortest_path_dist} shows the distribution of the shortest paths in ANI.

Table \ref{table:shortest_paths} summarizes the results of shortest paths analysis.
Apart from other statistics, we also show number of flights needed to be changed, which is an indicator of the convenience of travel in
the network; the lesser the better.
Around $99$\% of paths are reachable by changing a maximum of $2$ flights. Around $68$\% of paths are reachable by changing maximum
$1$ flight. Whereas about $7$\%  paths are connected by direct flights. We also deduce that the diameter (defined as the longest of all
shortest paths) of ANI is $4$, which means that one needs to change at most $3$ flights to reach from any airport to any other airport 
in ANI. 

From the passengers' point of view, best is as few change of flights as possible, to reach from any airport to any other airport in 
the network.
Though this is an ideal situation for the passengers, it is not economically viable for the air service  providers. 
We infer that the airport network has evolved, with the efforts from the service providers, to cater for the convenience of 
passengers, thus acquiring the small-world topology. 

\begin{table}
\caption{\label{table:shortest_paths}Break-up of the number of flight routes having a certain shortest path, their percentage 
and corresponding number of flights needed to be changed by the shortest route.}
\begin{ruledtabular}
\begin{tabular}{crrc} 
Shortest & No.\ of  & Percentage    & No.\ of Flights  \\
Path     & Paths    & of flight     &  needed to be     \\
         &          & routes        &  changed           \\
\hline
1 &   442 &   7.1720 & 0 \\
2 & 3741 & 60.7100 & 1 \\
3 & 1918 & 31.1260 & 2 \\
4 &     61 &   0.9899 & 3 \\ 
\end{tabular}
\end{ruledtabular}
\end{table}

\subsection{\label{subsec:degree_dist}Degree Distribution}
Degree, as defined in Eq.\/~\ref{eq:degree}, is one of the measures of centrality of a node in the network. 
Degree symbolizes the importance of a node in the network -- the larger the degree, the more important it is.
The distribution of degrees in a network is an important feature which reflects the topology of the network. It may shed
light on the process by which the network has come into existence. The networks in which the links between two
nodes are assigned randomly have a Poisson degree distribution~\cite{bollobas1981} with most of the nodes having a typical degree.
A large number of networks ranging from Internet to protein-protein interaction network in yeast, have been 
found~\cite{reka:thesis} to be having scale-free degree distributions. 
The scale-free distribution, characterized by a power law -- \mbox{$P(k) \sim k^{-\gamma}$}, 
with a scaling exponent $\gamma$ -- can be explained with the help of a growing network model with preferential attachment of the nodes 
which are being added to the network~\cite{SFNW:barabasi}.
 
\begin{figure}
\includegraphics{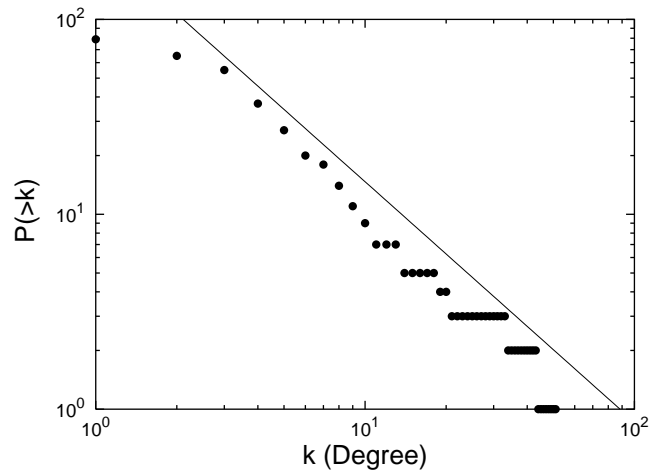}
\caption{\label{fig:cum_degree_dist}Cumulative degree distribution in ANI. It can be approximated by
the power-law fit, $P(k) \sim k^{-\gamma_{cum}}.$}
\end{figure}

Since ANI is a small network, we analyze the cumulative degree distribution, $P(>k)$, whose scaling exponent $\gamma_{cum}$ 
is related to that of $P(k)$ by $\gamma = \gamma_{cum}+1$. We find that the cumulative degree distribution of ANI follows a 
power law as seen in Fig.~\ref{fig:cum_degree_dist} for a wide range of $k$, although it deviates from it for 
large degrees. The deviation for large degrees can be attributed to the cost of adding links to nodes or to the limited capacity of 
nodes. With the help of numerical simulations, it was shown~\cite{air-amaral-PNAS} that the cost of adding links leads to a cutoff 
of the power-law regime for large $k$ in the cumulative degree distribution, as is the case with ANI.
We find the scaling exponent of ANI to be, $\gamma = 2.2 \pm 0.1.$ 

\section{\label{sec:wANI}Centrality in weighted ANI}
\begin{figure}
\includegraphics{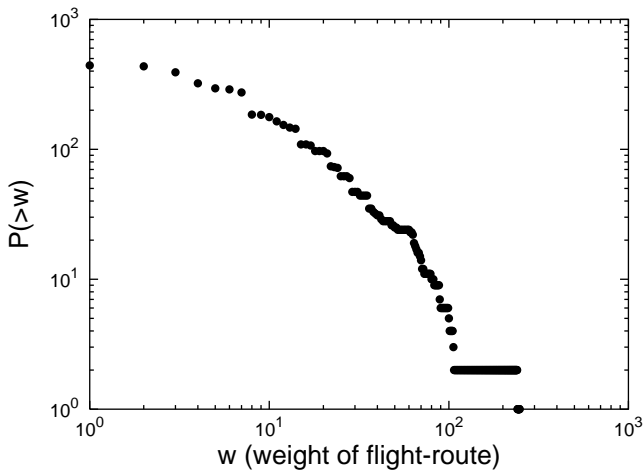}
\caption{\label{fig:cum_weights_dist}Cumulative weights distribution in ANI.}
\end{figure}

We analyzed the weighted ANI, by considering the flow of information (traffic) on the topology of the network. 
The statistical analysis of weights between pairs of nodes indicates the presence of right-skewed distribution as shown in 
Fig.~\ref{fig:cum_weights_dist}. This shows a high level of heterogeneity in ANI, as also found in the case of Airport Network
of China~\cite{air-china} and WAN~\cite{air-WAN-PNAS}. 

\begin{figure}
\includegraphics{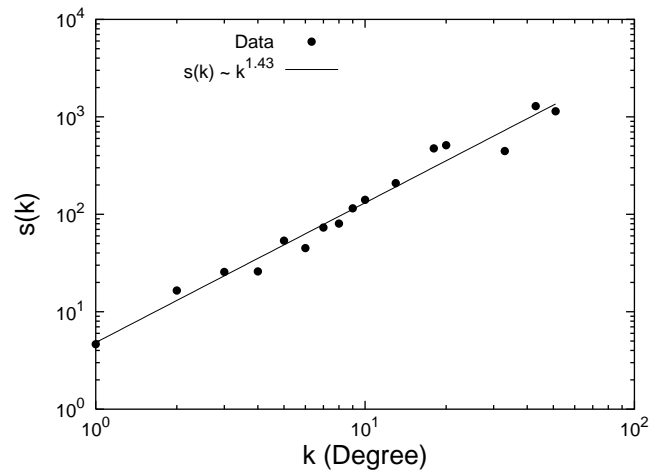}
\caption{\label{fig:sk}Average strength $s(k)$ as a function of degree (k) of nodes.}
\end{figure}

It has been observed that the individual weights do not provide a general picture of the network's complexity~\cite{yook:PRL}. 
We use a different measure of importance of a node which considers the flow of information in a network, following 
Barrat et al.~\cite{air-WAN-PNAS}. The weighted counterpart of degree, \emph{strength}~$(s_i)$, is defined as,
\begin{equation}
s_i = \sum_{j=1}^N a_{ij}w_{ij} .
\label{eq:strength}
\end{equation}
Strength of an airport represents total traffic handled by it per week. Fig.~~\ref{fig:sk} shows the correlation between degree ($k$)
and the average strength of vertices with degree $k$, $s(k)$. We find that $s(k)$ increases with $k$ as, $s(k) \sim k^{\beta}.$ 
If the strength and degree of a node were uncorrelated, then $s_i=\langle w \rangle k_i$, 
where $\langle w \rangle = (2M)^{-1} \sum_{i,j} a_{ij}w_{ij}$, which will yield $\beta=1.$ This is an uninteresting case, as in this 
situation weights do not provide any better information than degrees. 
For ANI, we find that the exponent to be $\beta_{ANI}=1.43 \pm 0.06.$ This implies that the strengths of nodes are strongly correlated
to their degrees in ANI. Larger is an airport, the more traffic it handles, in contrast to what would be expected if the topology and the
traffic were uncorrelated.  
This feature of ANI is similar to that of WAN, where it was found that $\beta_{WAN}=1.5 \pm 0.1.$

\section{\label{sec:anal_wANI}Structural organization in weighted ANI}
Various properties of network are decided, among many other things, by the way the network has evolved, the inherent 
interdependencies of the nodes, and the architectural constraints. Network parameters are defined so as to be able to capture  
such features of networks. We study few such parameters related to the local cohesiveness and connection tendency of nodes,
with and without consideration of weights to gain better insight of traffic dynamics on the architecture of  ANI. 

\subsection{\label{subsec:cc}Clustering Coefficient}
Clustering coefficient as defined in section~\ref{sec:uw-ani}, captures the local cohesiveness of a node. Average clustering 
coefficient ($C$) measures the global density of interconnected nodes in the network. Additional information at the intermediate 
level can be obtained by defining $C(k)$, average clustering coefficient of nodes with degree $k$. These topological parameters
overlook the flow of information on the network and hence may not present correct information about the network dynamics. 
The clustering coefficient could be redefined~\cite{air-WAN-PNAS}, to incorporate the weights of the edges, as,
\begin{equation}
c_{i}^{w} =   \frac{1}{s_i (k_i - 1)}  \sum_{j,h} \frac{w_{ij} + w_{ih} }{2} a_{ij}a_{ih}a_{jh}.
\label{eq:wt_cc}
\end{equation}
This parameter measures local cohesiveness, by taking into account the interaction intensity found on the local triplets.
The unweighted and weighted clustering coefficients could be compared (the normalization ensures that $0 \leq c_{i}^{w} \leq 1$)
to assess the tendency of accumulation of traffic on interconnected triplets.  
Weighted clustering coefficients averaged over all nodes ($C^w$) and over all nodes with degree $k$ ($C^w(k)$) are defined 
analogous to their topological counterparts.

It has been shown~\cite{ravsaz:PRE} that hierarchical networks are expected to have a non-trivial, power-law decay of $C(k)$ as a 
function of $k$, which means low degree nodes belong to interconnected communities. 
Many real networks have been found~\cite{ravsaz:PRE,ravsaz:science,vazquez:PRE} to be having such nontrivial decay. 
We find that in ANI, $C(k)$ shows a power-law decay with $C(k) \sim k^{-1}.$ This points toward an inherent hierarchy in 
the architecture of ANI. This is an interesting result as it has been found~\cite{ravsaz:PRE} that some other infrastructure 
networks (Internet at the router level and power grid network of Western United States) which are constrained by 
``geographical organization'' do not show such hierarchy. This could be because, as opposed to the above-mentioned 
networks, where the cost of having a link grows in proportion to the distance between two nodes, the geographical constraint is 
not so strong in the case of airport networks. It is relatively easier to add an air connection to the existing network, than to add
a data cable or an electrical cable for comparable geographical distances.

\begin{figure}
\includegraphics{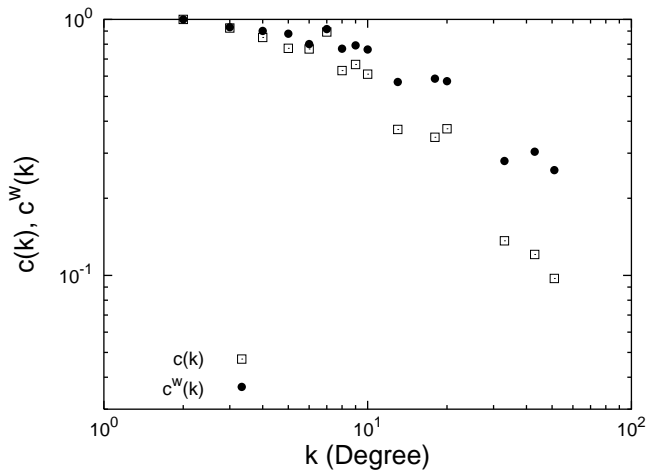}
\caption{\label{fig:cc}Average unweighted ($C(k)$) and weighted ($C^{w}(k)$) clustering coefficients of nodes with degree $k$.}
\end{figure}

As seen in Fig.~\ref{fig:cc}, after remaining almost constant for $k<8$, $C(k)$ falls rapidly thereafter. 
This indicates that large airports provide air connectivity to far-off airports, which themselves do not tend to be connected. 
In ANI, regional and national hubs provide air connectivity to airports in their domains, with the latter providing it to a much larger 
number, which reduces the $C(k)$ of these hubs. We find that $C^w/C \cong 1.075$, which means that the traffic is accumulated on 
interconnected groups of nodes which form high traffic corridors, also known as \emph{trunk routes}. 
We find that $C^w(k)$ is restricted in its range across $k$ and is consistently higher than the corresponding $C(k).$ 
The higher the $k$, more pronounced is the difference between $C^w(k)$ and $C(k)$. 
This implies that with increasing degree, airports have progressive tendency to form interconnected groups with high-traffic links. 
This \emph{rich-club phenomena}~\cite{rich-club-phenomena}, in which high degree nodes tend to form cliques with 
nodes with equal or higher degree, is also observed in WAN~\cite{air-WAN-PNAS}.

\subsection{\label{subsec:degree_corr}Degree Correlations}
Another parameter which is used to investigate the networks' architecture is the average degree of nearest neighbors, $k_{nn}(k)$,
for nodes of degree $k$~\cite{satorras:knn}. It is convenient to calculate $k_{nn}(k)$, which is related to the correlations between
the degrees of connected vertices as $k_{nn}(k) = \sum_{k'} k' P(k'|k),$ where $P(k'|k)$ is the conditional probability that a given 
vertex $k$ is connected to a vertex of degree $k'.$ For a topology with no correlations among nodes' connectivity, the conditional
probability as well as $k_{nn}(k)$ are independent of $k.$ On the contrary, explicit dependence on $k$ is a signature of non-trivial 
correlations among the nodes' connectivity and the possible presence of a hierarchical structure in the network 
topology. In the presence of correlations, a network can be classified as one having ``assortative mixing'' or
``disassortative mixing'', based on whether $k_{nn}(k)$ is an increasing or decreasing function of $k$, respectively. In the former
class, high degree nodes have tendency to be connected to nodes with large degrees, whereas in the latter class high degree nodes
have a majority of low degree nodes as their neighbors.
For a weighted network the \emph{weighted average nearest neighbors degree}~\cite{air-WAN-PNAS} could be defined as,
\begin{equation}
k_{nn,i}^w = \frac{1}{s_i} \sum_{j=1}^N a_{ij}w_{ij}k_{j},
\label{eq:knnw}
\end{equation}
which measures the effective \emph{affinity} to connect with high- or low-degree neighbors according to the magnitude of the 
actual interactions. The trend of $k_{nn}^w$ indicates the weighted assortative or disassortative nature of the network
considering the strength of interactions among the nodes.

\begin{figure}
\includegraphics{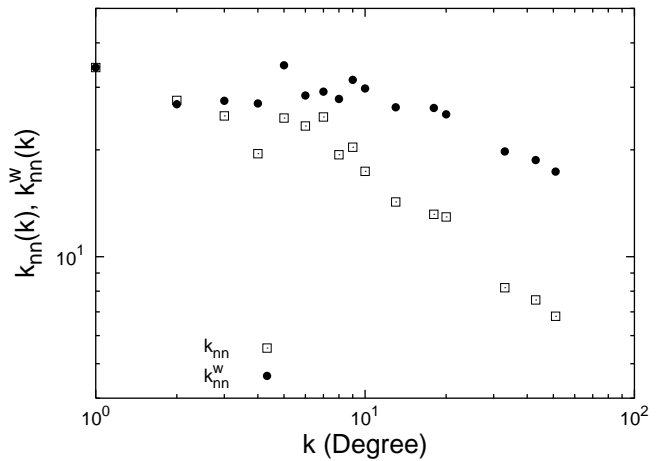}
\caption{\label{fig:degree_corr}Average unweighted ($k_{nn}(k)$) and weighted ($k_{nn}^w(k)$) degree of nearest neighbors of 
nodes with degree $k$.}
\end{figure}

We find (Fig.~\ref{fig:degree_corr}) that ANI has uncorrelated structure for small degrees ($k<8$) with no definite assortative
or disassortative mixing. But for larger degrees ($k \geq 8$) ANI shows a clear disassortative mixing. This property of ANI is 
drastically different from that of WAN which was found to be having assortative mixing for small degrees ($k<10$) beyond which it 
approaches a constant value. The consistent disassortativity in ANI for higher degrees could be attributed to the political 
compulsions on regional and national hubs to provide connectivity to a large number of low-degree destinations. 
A hub in WAN is connected to many other hubs (in other countries), thus compensating for the large number of connections with 
low-degree airports. Since the number of hubs is limited in a national airport network, such as ANI, it could lead to disassortative
topology.
This feature of ANI is consistent with the nature of Airport Network of China~\cite{air-china} which is reported to be disassortative.
This could well be a feature of national infrastructure networks which have similar geo-political compulsions. 
The normalized correlation function ($r$)~\cite{r:newman} ($-1 \leq r \leq 1$) is another global quantitative measure of degree 
correlations in a network, which is zero for no correlations among nodes' connectivity and positive or negative for assortative 
or disassortative mixing, respectively. For ANI we find $r=-0.4016$, indicating disassortative mixing which is 
consistent with the observation in Fig.~\ref{fig:degree_corr}. Many technological networks and biological networks are
found to be disassortative, while social networks are found to be assortative~\cite{r:newman}. 
Indian Railway Network~\cite{transport:manna} has also been found to be disassortative.

We find that weighted degree correlation, $k_{nn}^w(k)$, is restricted in its range across $k$ and is consistently higher than 
its unweighted counterpart, $k_{nn}(k)$, indicating toward bias in the traffic in ANI. 
The larger the degree of a node, the more pronounced is the difference. This implies that despite disassortative topology, the traffic is 
concentrated between high degree nodes. This is understandable as most airline service providers concentrate on certain 
profitable, so called, \emph{trunk routes}, thereby creating high traffic corridors~\cite{report:bw}.

\section{Conclusions}
\label{sec:conclusions}
We find that ANI, despite being small in size, has complex dynamics similar to those of bigger air transportation networks.
It represents an evolving transport infrastructure of a developing nation.
It presents a case of a national airport network which has a different mechanism of formation than the global network.
ANI, whose topology has a signature of hierarchy, has small-world network features and is characterized by a truncated 
scale-free degree distribution.  
Analysis of weighted ANI reveals clearer picture of the network dynamics.
The traffic in ANI is found to be accumulated on interconnected groups of airports. 
It has disassortative mixing in contrast to WAN, with hubs having large number of low-degree neighbors. 
The topological disassortativity is offset by the traffic dynamics as the traffic is concentrated between high-degree nodes.

ANI is expected to grow at a rapid speed with addition of airports, several low-cost air services~\cite{report:bw}, 
and importantly by increase in strength and complexity of interactions. 
It will be interesting to study the evolution of this air transportation network, as it offers
an excellent example to understand and model the topology, traffic dynamics and geo-political constraints which shape 
a national infrastructure of economic importance.

\begin{acknowledgments}
The author is grateful to Somdatta Sinha, Marc Barth\'{e}lemy, and S S Manna for valuable discussions, and Madhavi for help
in obtaining data. 
The author is thankful to Deepak Dhar, R E Amritkar, V Suresh, and P K Mohanty for their comments on the manuscript.
Author thanks Council of Scientific and Industrial Research (CSIR), Govt.\ of India, for the Senior Research Fellowship. 
\end{acknowledgments} 


\end{document}